\documentclass[aps,prl,twocolumn,floatfix]{revtex4-1}

\usepackage{graphicx}
\usepackage{amssymb}
\usepackage{amsmath}
\usepackage{epstopdf}
\usepackage{hyperref}
\usepackage{cleveref}

\crefname{equation}{Eq.}{Eqs.}
\crefname{figure}{Fig.}{Figs.}
\crefname{appsec}{Appendix}{Appendices}

\newcommand{\nn}{\nonumber}

\newcommand{\veps}{\varepsilon}

\begin{document}

\title{Reversible Pore Gating in Aqueous Mixtures via External 
Potential}

\author{Sela Samin}

\affiliation{Institute for Theoretical Physics, Center for Extreme Matter and 
Emergent Phenomena, 
Utrecht University, Princetonplein 5, 3584 CC Utrecht, The Netherlands}

\author{Yoav Tsori}
\affiliation{Department of Chemical Engineering and the Ilse Katz Institute for 
Nanoscale
Science and Technology, Ben-Gurion University of the Negev, 84105 Beer-Sheva, 
Israel.}

\date{April 20, 2016}

\begin{abstract}
We examine theoretically porous membranes in aqueous mixtures. 
We show that large membrane pores can be reversibly gated 
from `off' (co-solvent-rich, poor conductor of ions and other solutes) to 'on' 
(water-rich, good conductor) states by an external potential. The transition voltage 
or charge for 
switching depends on the membrane hydrophilicity/hydrophobicity, the salt content, the 
preferential solvation of the salt ions, and the temperature. These parameters also 
determine whether the filling transition is abrupt or gradual.
\end{abstract}

\maketitle

Transport of liquids and solutes across nano- and meso-scale pores occurs in 
many biological and synthetic membranes \cite{katchalsky_bpa_1958,hoek_ees_2011}. 
Considerable 
research on the pore chemistry and shape has been carried out to improve the membrane's 
conductivity and selectivity and to prevent the membrane from clogging 
\cite{neil_science_2013}. In pore gating, electric fields have advantages over 
the use of thermal variations \cite{yang_jms_2003,azzaroni_small_2009} 
or pressure difference across the membrane \cite{aizenberg_nature_2015} since they can be 
easily switched on and off and do not require the membrane to be mechanically robust. 
Current works consider electric fields that are parallel to the pore's axis and pure water 
or coexistence of vapor and water 
\cite{hansen_jcp_2004,hansen_jcp_2005,siwy_nature_nanotech_2011,lavrik_acsnano_2011}. 

Here 
we propose a novel methodology to reversibly gate hydrophobic or hydrophilic pores by 
using aqueous solutions and fields that are perpendicular to the pore's walls. 
For hydrophobic membranes, the pore opens and fills with water when voltage is 
applied to the membrane. The pore closes by filling with the co-solvent by 
natural diffusion on a time that scales as the pore size squared once the 
voltage is removed. Ionic current ceases a short time after turning on or off 
the voltage. 

The pore filling described below is promoted by the preferential solvation of ions in 
water, which 
has previously been shown to modify bulk coexistence 
\cite{tsori_pnas_2007,onuki_pre_2010,bier2012}, solvent and ion adsorption on 
surfaces \cite{tsori_pnas_2007, onuki_pre_2010,bier2012}, and the 
inter-particle potential in colloidal 
suspensions \cite{leunissen2007,nellen2011,samin2014,efips_epl,
onuki_pre_2011, efips_jcp_2012, pousaneh2014}. Here, a purely solvation induced 
transition in confinement is quantified for the first time, and it is shown that 
its magnitude is comparable to that of capillary condensation. More importantly, 
the filling transition is predicted to occur even for highly hydrophobic pores, 
unlike capillary condensation. Even more so, continuous filling by an external 
potential is predicted for hydrophobic pores far above the mixture critical 
temperature.

We consider a small pore in a membrane embedded in a large reservoir of aqueous mixture.
For large membrane potentials the ion density at the surfaces becomes very high and 
therefore we use a Modified Poisson-Boltzmann (MPB) approach \cite{borukhov1997} 
employing the incompressibility constraint $\phi+\phi_{\rm cs}+v_0 n^-+v_0 n^+=1$, where 
$\phi$ and $\phi_{\rm cs}$ are the volume fractions of water and 
cosolvent respectively, $v_0$ is the common molecular volume of all species, 
and $n^\pm$ is the number density of cations and anions, respectively.

The solvent free energy density of mixing is
\begin{align}
\label{eq:fmix}
f_{\rm m}&=\frac{k_BT}{v_0}\left[
\phi\log(\phi)+\phi_{\rm cs}\log(\phi_{\rm cs})+\chi\phi \phi_{\rm cs}\right]~.
\end{align}
$k_BT$ is the thermal energy and 
$\chi\sim1/T$ is the Flory interaction parameter. 
\cref{eq:fmix} leads to an Upper Critical Solution Temperature type phase diagram.
In the $\phi-T$ plane, a homogeneous phase is stable above the binodal curve, 
$\phi_b(T)$, whereas below it 
the mixture separates to water-rich and water-poor phases with compositions 
given by $\phi_b(T)$. The two phases become indistinguishable at the critical 
point $(\phi_c,\chi_c)=(1/2,2)$.

The free energy density of the ions dissolved in the mixture, $f_{\rm i}$, 
is modeled as
\begin{align}
\label{eq:fions}
 f_{\rm i}=&k_BT\bigl[n^+\log (v_0n^+)+n^-\log 
(v_0n^-) \nn \\
&-\phi(\Delta u^+n^+ + \Delta u^-n^-)\bigr],
\end{align}
The first line in \cref{eq:fions} is the entropy of the ions and the second 
line models the solvent specific short-range interactions between ions and 
solvents, 
where the solvation parameters, $\Delta u^\pm$, measure the preference of ions 
towards a local water environment 
\cite{onuki_jcp_2004,tsori_pnas_2007,onuki_pre_2011}. We choose a simple linear 
solvation model for clarity. Nonetheless, we note that our 
results remain qualitatively the same and quantitatively similar when more 
complex solvation models are employed, such as the one offered by Bier 
\textit{et al.} \cite{bier2012}.

Experiments show that $\Delta G_t$, the Gibbs transfer energy for 
transferring an ion from a solvent with composition $\phi_1$ to a solvent with 
composition $\phi_2$ is on the order of $1-10k_BT$ in aqueous mixtures 
\cite{marcus_cation,*marcus_anion} and since $\Delta G_t=k_B 
T\Delta u^\pm(\phi_2-\phi_1)$ one finds that $\Delta u^\pm \sim 1-10$.
In general $\Delta G_t$ is highly ion specific and also depends strongly on the ion sign 
and valency. The filling transition is robust; for clarity of presentation we 
present only the simple case where $\Delta u^+=\Delta u^-$.

The electrostatic energy density, $f_{\rm e}$, for a monovalent salt is given by
\begin{align}
\label{eq:fes}
 f_{\rm e}=-\frac{1}{2}\veps(\phi)(\nabla \psi)^2+e(n^+-n^-)\psi
\end{align}
where $\psi$ is the electrostatic potential, $e$ is the elementary charge and 
$\veps(\phi)$ is the permittivity, 
assumed to depend linearly on the mixture composition: 
$\veps(\phi)=\veps_{\rm cs}+(\veps_w-\veps_{\rm cs})\phi$, where $\veps_w$ and 
$\veps_{\rm cs}$ 
are the water and co-solvent permittivities, respectively. This $\phi$ 
dependence of $\veps$ leads in nonuniform electric fields to a 
dielectrophoretic force which attracts the high permittivity water towards 
charged surfaces \cite{efips_jcp_2012,tsori_rmp_2009}.

The surface energy density $f_s$ due to the 
contact of the mixture with a solid surface is given by:
\begin{equation}
f_s=k_BT\Delta\gamma\phi({\bf r}_s)+e\sigma\psi({\bf r}_s),
\label{eq:fsions}
\end{equation}
where ${\bf r}_s$ is a vector on the surface. The first term in 
\cref{eq:fsions} models the short-range interaction between the fluid and the 
solid. The parameter $\Delta\gamma$ measures the difference between the 
solid-water and solid-cosolvent surface tensions. The second term in 
\cref{eq:fsions} is the 
electrostatic energy when the surface carries a charge density $e\sigma$.

\begin{figure}[t]
\includegraphics[width=3.5in,clip]{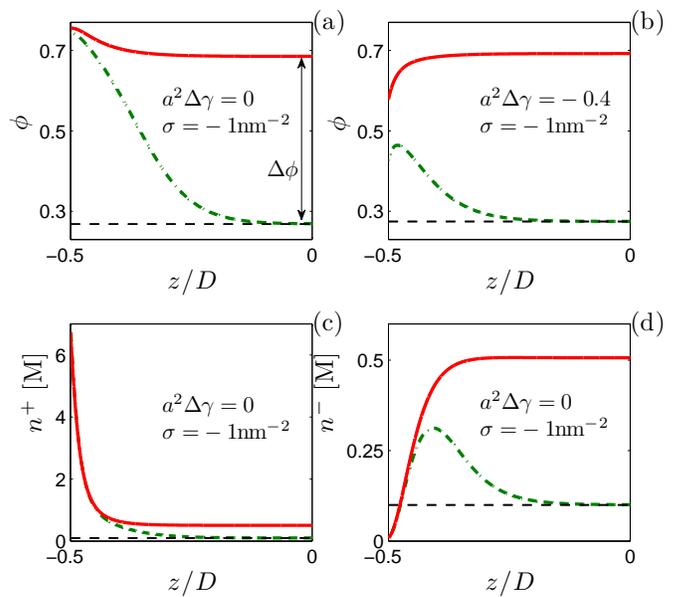}
\caption{Composition and ion density profiles for bulk compositions $\phi_0$ 
just before (dash-dot lines) and after (solid lines) the filling composition 
$\phi^*$. (a) 
Composition profiles for a charged and chemically indifferent pore (see 
figure for parameter values).  
The composition ``jump'' at the midplane of the pore is denoted by $\Delta 
\phi$. (b) A charged hydrophobic pore. 
(c-d) the cation and anion density profiles corresponding to (a). The 
dashed lines in (a)-(b) correspond to $\phi_0$ and in (c)-(d) to 
$n_0$. We used $t = 0.96$, $n_0 = 0.1$ M. Unless otherwise stated, in all 
figures we 
took $D =10$ nm, $\Delta u^\pm=4$, $\veps_w=79.5$, $\veps_{\rm cs}=6.9$ and 
$T_c=307$ K. $a=v_0^{1/3}=3.4$ \AA{} is 
a molecular length.}
\label{fig_prof}
\end{figure}

The pore is modeled as two identical 
flat plates with area $A$ located at $z=\pm D/2$ (pore width is $D$). The pore walls can 
either
carry a fixed surface charge density $e\sigma$ or have a constant potential 
$\psi_s$. The pore is in contact with a reservoir for which the 
electrostatic potential is $\psi=0$. The pore walls can be hydrophilic ($\Delta 
\gamma>0$) of hydrophobic $(\Delta \gamma<0)$. In this planar geometry, the 
grand potential is given by 
\begin{align}
\label{eq:Omega}
&\Omega[\phi(z),n^\pm(z),\psi(z)]=A\int\biggl[\tfrac{1}{2}C\left(\nabla 
\phi\right)^2
+f_{\rm m}+f_{\rm i}+f_{\rm e} \nn \\ 
&-\lambda^+n^+-\lambda^-n^--\mu 
\phi/v_0\biggr]{\rm d} z + 2A f_s. 
\end{align}
The square-gradient term accounts for 
the energetic cost of composition inhomogeneities, 
where $C=k_BT\chi/v_0^{1/3}$ is a positive constant \cite{efips_epl}. 
The chemical potentials of the cations and anions are $\lambda^\pm$, respectively, and 
that of the mixture is $\mu$.

The pore's behavior follows from the profiles $\phi(z)$, $\psi(x)$, and $n^\pm(x)$. The 
variational equations $\partial\Omega/\partial n^\pm=0$ allow isolation of 
$n^\pm$ in the form of a Boltzmann-like distribution as a function of $\phi$ 
and $\psi$ (see 
Supplementary material). We 
solve numerically the two other equations $\partial\Omega/\partial \phi=0$ and 
Gauss's 
law $\nabla \cdot (\veps(\phi)\nabla \psi)=e(n^--n^+)$. The boundary conditions for 
$\phi$  at the walls are ${\rm d}\phi/{\rm d}z=\Delta \gamma/C$ 
\cite{andelman_jpcb_2009}. The condition for the potential is either ${\rm 
d}\psi/{\rm d}z=-e\sigma/\veps(\phi)$ or 
$\psi=\psi_s$, corresponding to fixed charge density or potential, respectively.

We focus on homogeneous water-poor mixtures with a small enough average bulk 
composition $\phi_0$: 
$\phi_0<\phi_t<\phi_b$. $\phi_t$ is the composition for which a bulk mixture 
first undergoes a so-called precipitation transition \cite{onuki_pre_2010}, 
where small water-rich droplet begin to nucleate due to the preferential 
solvation of ions in them. The derivation of $\phi_t$ is detailed in 
the Supplementary material. We find that in the vicinity of 
$\phi_t$ two types 
of configurations are possible in equilibrium, distinguished by the value of the 
composition at the pore midplane, 
$\phi(z=0)$: either the pore has the bulk composition 
$\phi(z=0)\approx\phi_0<\phi_c$ or it fills  
with water and then $\phi(z=0)>\phi_c$. 
The stable configuration is the one for which the grand potential is lower. 
A first-order filling transition from one configuration to the other  
occurs when the corresponding grand potentials are equal at a composition $\phi_0=\phi^*$.

The composition and ion density profiles in the vicinity of the filling 
transition are shown in \cref{fig_prof} for several scenarios. 
In this figure only the region $-D/2\le z\le0$ of the symmetric profiles is shown.
The reduced temperature is set to $t\equiv T/T_c=0.96$, where $T_c$ is the 
mixture 
critical temperature, and the salt is assumed 
hydrophilic, $\Delta u ^\pm=4$. Dash-dot and solid curves correspond to profiles just 
before and after the transition, respectively, while 
the horizontal dashed lines show the bulk values. A purely 
solvation-induced filling transition is demonstrated in 
\cref{fig_prof} (a), where the pore walls are chemically indifferent, 
$\Delta\gamma=0$, but are highly charged. Here, prior to the transition, a 
wide adsorption layer is created when the counter-ions ``drag'' with them 
the water to the walls such that $\phi(z=-D/2)>\phi_c=1/2$. 
The thickness of the 
layer here is associated with a modified Debye length 
$\lambda_D(n_0;\phi_0,\Delta u)$ (For more details see Ref. \cite{efips_epl}). 
At the transition, the composition profile jumps discontinuously to 
$\phi(z)>\phi_c$ throughout the pore volume. 

The corresponding ion profiles in \cref{fig_prof} (c)-(d) show that 
the ion density decays to its bulk value closer to the wall than the 
composition because composition gradients are energetically costly.
Therefore, farther from the wall the relevant length scale is a 
modified bulk correlation length, $\xi(T;n_0,\Delta u)$, associated 
with the width of interfaces and depending strongly on $T-T_c$ 
\cite{efips_epl}. The co-ions are electrostatically repelled from the wall but  
are also drawn to it due to their favorable solvation in water, resulting 
in a maximum in the co-ions profile, shown in \cref{fig_prof} (d). 
Close enough to the binodal it is energetically favorable to eliminate the 
large composition gradients and the filling transition ensues. We stress that once filling 
takes place the profiles do not decay to the bulk values far from the wall.

The filling transition is predicted to occur even for a charged but highly hydrophobic 
pore 
with $a^2\Delta\gamma=-0.4$. \cref{fig_prof} (b) shows the composition 
profiles for this case, where although the solvent is depleted close to the 
wall (evidenced by the positive slope $\phi'(z=-D/2)$), the attraction of the 
counter-ions 
to the walls together with the preferential solvation leads to an increase  
in the water composition and eventually to the filling of the pore. 
Although the short-range interaction of the solvent with the pore 
walls can also promote filling as in regular capillary condensation, here the importance 
of selective solvation, which is a volume contribution to the free energy, is much larger.
This holds when the pore walls are highly charged, the salt concentration 
is large and the coupling with the solvent is strong.

\begin{figure}[!tb]
\includegraphics[width=3.3in,clip]{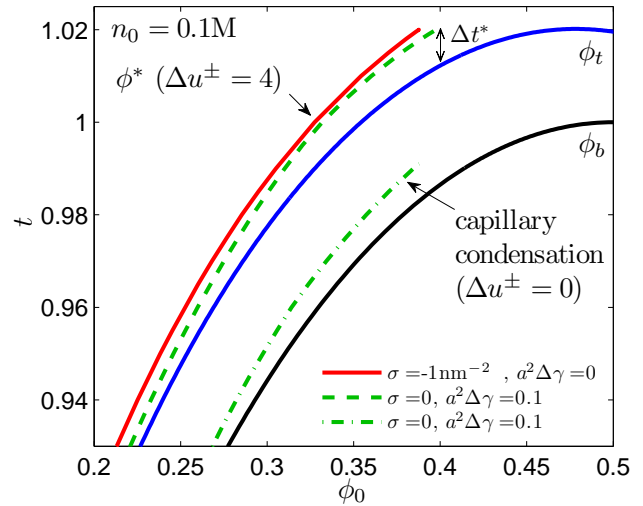}
\caption{The filling composition curve $\phi^*$ vs reduced 
temperature $t=T/T_c$ for membranes with pores of width $D=10$ nm embedded in 
aqueous 
mixtures with hydrophilic ions with $\Delta u^\pm=4$. The finite volume of the ions is
neglected.
Topmost solid line (red): $\phi^*$ for a charged and chemically indifferent pore 
($\sigma=-1$nm$^{-2}$, $\Delta\gamma=0 a^{-2}$). The filling effect is 
weaker when the pore is uncharged but hydrophilic ($\sigma=0$, 
$\Delta\gamma=0.1 a^{-2}$), dashed line. Solid curve denoted as $\phi_t$ is the 
mixing-demixing phase boundary in the bulk. 
The regular capillary 
condensation line for hydrophilic pores ($\Delta u^\pm=0$, $\Delta\gamma=0.1 a^{-2}$) 
is plotted as a reference (dash-dot line). $\phi_b$ is the bulk mixing-demixing binodal 
line in the absence of salt.}
\label{fig_bin}
\end{figure}

The stability of the mixture in the pore is determined by the filling curve 
$\phi^*(T)$ in the $\phi_0-T$ plane.
This curve is the value of the bulk composition $\phi_0$ for which filling 
occurs, 
for different temperatures. The filling curve for $D=10$ nm, $n_0=0.1$ M and 
$\Delta u^\pm=4$
is shown in \cref{fig_bin}. We plot $\phi^*(T)$ 
for a highly charged and chemically indifferent pore (solid, 
$a^2\Delta\gamma=0$, $\sigma=-1$nm$^2$) and an uncharged and hydrophilic pore 
(dashed, $a^2\Delta\gamma=0.1$, $\sigma=0$). For comparison, we also 
plot in 
\cref{fig_bin} the capillary condensation curve of a salt-free mixture 
with hydrophilic walls (dash-dot, $a^2\Delta\gamma=0.1$). In all cases, the 
discontinuity of the physical quantities across the curve vanishes at a film 
critical point and increase with temperature decreasing away from it. 

The temperature shift of the reciprocal filling curve 
$T^*(\phi_0)$ under confinement, $\Delta t^*$, is defined as
\begin{equation}
\Delta t^*=t^*(\phi_0;D)-t^*(\phi_0;D\rightarrow\infty)=t^*(\phi_0;D)-t(\phi_t)~,
\end{equation}
where $\phi_t(t)$ is the bulk mixing-demixing curve.
$\Delta t^*$ quantifies the magnitude of the filling phenomena and should be 
readily accessible experimentally.

\begin{figure}[!tb]
\includegraphics[width=3.3in,clip]{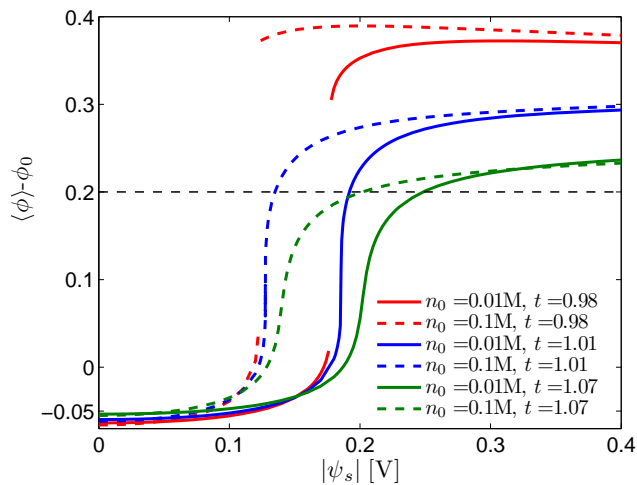}
\caption{Pore 
filling with external potentials. Curves show the difference between the 
average pore water fraction $\langle\phi\rangle$ and the bulk value $\phi_0$ vs pore
potential $|\psi_s|$ for different salt content 
and temperatures. The filling is discontinuous for $t=0.98$ and gradual 
for $t=1.01$ and $t=1.07$. At zero or small voltages, $\langle\phi\rangle$ is 
smaller than the 
bulk value since the pore is hydrophobic ($a^2\Delta \gamma=-0.4$). For 
voltages 
below $\sim -0.22$ V water becomes the majority component in the pore, 
$\phi>0.5$ 
($n_0=0.01$ M). For $n_0=0.1$ M mixtures this occurs at smaller value of 
$|\psi_s|$.
The bulk composition is $\phi_0=0.3$ and $D=5$ nm.
}
\label{fig3}
\end{figure}

Based on the above insight we now turn to check whether indeed pore filling can be
gated by a change of the external potential. Tuning of the potential can be 
realized by the use of activated carbon based membranes, as in e.g. supercapacitors.
We find that the average water fraction $\langle\phi\rangle$ in a hydrophobic pore 
increases with increasing (negative) wall potential $|\psi_s|$ even when there is 
no discontinuous filling transition, see \cref{fig3}. At zero potential, 
$\langle\phi\rangle$ is negative as water is depleted from the pore. 
Clearly for $|\psi_s|>0.22$ V ($n_0=0.01$ M) or for $|\psi_s|>0.15$ V 
($n_0=0.1$ M), water is a 
majority in the pore. 

\begin{figure}[!tb]
\includegraphics[width=3.3in,clip]{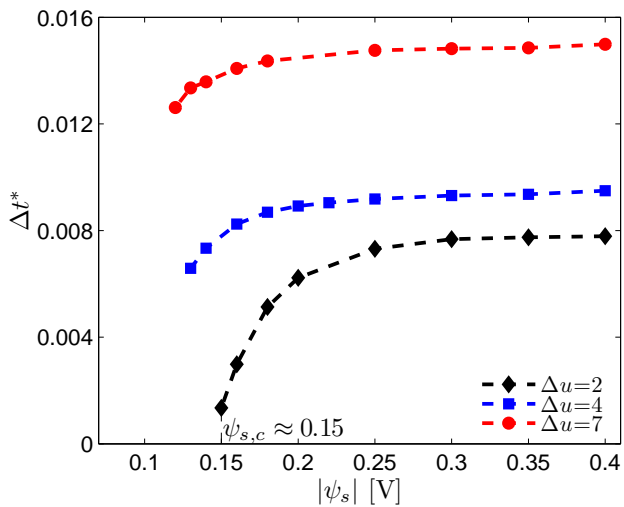}
\caption{The temperature shift $\Delta t^*$ as a 
function of the wall
potential $\psi_s$ for a hydrophobic pore and different values of
$\Delta u^\pm$. Here $t=0.96$,
$a^2\Delta\gamma=-0.4$ and $D=10$ nm. Lines are a guide to 
the eye.}
\label{fig_V}
\end{figure}

In \cref{fig_V} we plot $\Delta t^*$ vs the surface 
potential $\psi_s$ for a highly hydrophobic pore and for several values of 
$\Delta u$. Since the pore is hydrophobic the 
filling occurs only above a critical value for the potential and the 
magnitude of $\Delta t^*$ strongly depends on $\Delta u$. The value of the critical 
voltage (the value of $|\psi_s|$ of the bottom-left point of each curve) decreases with 
increasing $\Delta u$, although not dramatically. 
Above the critical voltage $\Delta t^*$ increases with $|\psi_s|$ until it 
saturates. This saturation occurs more quickly with large ionic selectivity (large values 
of $\Delta u$).

In summary, we use a simple modified Poisson-Boltzmann theory to describe the 
thermodynamics of porous 
membranes immersed in aqueous mixtures. We show that pore filling transitions 
are influenced by the confinement of the pore and occur at temperatures 
significantly above the so-called precipitation temperatures of mixtures with ions 
without the pore confinement. The theory can be used to predict 
whether the pore is filled with the cosolvent-rich or with the water-rich 
phases. 

\cref{fig_V} and \cref{fig3} show abrupt and gradual filling transitions with 
increasing pore potential, respectively; equivalently this means sensitivity to the 
pore's surface charge. For membranes with immobile charges the surface heterogeneity
leads to variations of the surface charge. In such membranes some pores will be 
``open'' while some will be ``closed'' and the conductance of liquids or solutes across 
the membrane will be heterogeneous accordingly.

For composite porous carbonaceous membranes and other types of solid membranes, 
\cref{fig3} and \cref{fig_V} demonstrate the feasibility 
of robustly and sensitively filling or emptying the pore with water in an 
on/off manner by 
connecting the membrane to an external potential. This filling will lead to 
reversible blockage or clearance of the membrane to small solutes, depending on their 
hydrophobicity or hydrophilicity. 

The continuous filling of 
a very hydrophobic pore at temperatures up to $20$ K above $T_c$ in 
\cref{fig3} suggests that our predictions could be tested, for example, 
using water-acetonitrile at room temperature.

S. S. acknowledges funding from the European Union’s Horizon 2020 research and 
innovation programme under the Marie Sk\l{}odowska-Curie grant agreement No. 
656327. Y. T. acknowledges the support from COST Action
MP1106 and Israel Science Foundation Grant No. 56/14.

\appendix
\crefalias{section}{appsec}
\renewcommand{\theequation}{S.\arabic{equation}}
\setcounter{equation}{0}

\section{Supplementary material}
\label{sec_mpb}

\noindent {\it Profiles}. In equilibrium the profiles $\phi(z)$, $\psi(z)$ and 
$n^\pm(z)$ are the extremizers of 
the grand potential, and thus they obey
\begin{widetext}
\begin{align}
\label{eq:compmpb}
\frac{\delta \Omega}{\delta \phi}=&-C
\nabla^2\phi+\frac{k_BT}{v_0}\biggl[\log
\left(\frac{\phi}{1-v_0n^+-v_0n^--\phi}\right)+\chi(1-2\phi-v_0n^+-v_0n^-)\biggr
] \nn
\\
&-\frac{1}{2}
\frac{d\veps}{d\phi}(\nabla\psi)^2-k_BT(\Delta
u^+n^++\Delta u^-n^-)-\mu/v_0=0~,\\
\label{eq:mpb}
\frac{\delta \Omega}{\delta \psi}=&\nabla \cdot (\veps(\phi)\nabla
\psi)+(n^+-n^-)e=0~,\\
\label{eq:ionmpb}
\frac{\delta \Omega}{\delta n^\pm}=&\pm e \psi+k_BT\biggl[\log
\left(\frac{v_0n^\pm}{1-v_0n^+-v_0n^--\phi}\right)-(\Delta
u^\pm+\chi)\phi\bigg]-\lambda^\pm=0~.
\end{align}
\end{widetext}

$n^\pm$ in \cref{eq:ionmpb} can be isolated to give
\begin{align}
\label{eq:npm_exp1}
n^\pm=\frac{P^\pm (1-\phi)}{v_0(1+P^++P^-)}~,
\end{align}
where $P^\pm$ is given by
\begin{align}
P^\pm=\exp\left(\frac{\mp e \psi}{k_BT}+(\Delta
u^\pm+\chi)\phi+\frac{\lambda^\pm}{k_BT}\right)~.
\end{align}
One can put \cref{eq:npm_exp1} in
\cref{eq:mpb,eq:compmpb} and solve to obtain the electrostatic potential and 
composition profiles. From these profiles we determine all other quantities. 
In the Poisson-Boltzmann limit of point-like ions, $v_0n^+,v_0n^-\rightarrow 
0$, \cref{eq:compmpb} reduces to %
\begin{align}
\label{eq:comp}
&-C
\nabla^2\phi+\frac{\partial f_m}{\partial \phi}-\frac{1}{2}
\frac{{\rm d }\veps}{{\rm d }\phi}(\nabla\psi)^2 
\\ \nonumber &-k_BT\left[\Delta
u^+n^+ + \Delta u^-n^-\right]-\mu/v_0=0,
\end{align}
and the ion densities follow the Boltzmann distributions
\begin{align}
\label{eq:npm}
n^\pm&=n_0 e^{\mp e \psi/k_BT+\Delta u^\pm(\phi-\phi_0)}.
\end{align}

\noindent {\it Determination of the composition $\phi_t$}. Consider a small 
arbitrary volume of the mixture within a homogeneous bulk with 
a composition $\phi_0$ and a dissolved salt of point-like ions 
with a concentration $n_0$. In the absence of external electric
fields ($\psi=0$) or surface fields ($a^2\Delta \gamma =0$) the composition and 
ion densities in this volume are also homogeneous. The total grand density in 
this case, $\omega_0$, is written as
\begin{align}
\label{eq:tom}
\omega_0&=f_0-\lambda^+n^++\lambda^-n^- -\mu\phi/v_0.
\end{align}
where $f_0$=$f_{\rm m}+f_{\rm i}$ and $\lambda^\pm$ and $\mu$ are the chemical 
potentials of the ions and water, respectively. $f_{\rm i}$ 
defined in \cref{eq:fions} of the main text. In the limit of point-like ions, 
$f_{\rm m}$ 
from  equation \cref{eq:fmix} of the main text reduces to 
\begin{align}
\label{eq:fmpb}
f_{\rm m}=\frac{k_BT}{v_0}\left[\phi\log(\phi)+(1-\phi) \log(1-\phi)+ 
\chi\phi(1-\phi)\right] 
\end{align}

For hydrophilic ions, $\Delta u^\pm>0$, and a 
water-poor reservoir, $\phi_0<1/2$, $\omega_0$ is 
modified such that a water-rich phase in the small volume may coexist 
with the water-poor phase of the 
reservoir. The composition and salt
concentration of each phase are determined by the equality of
chemical potentials in the small volume and reservoir:
\begin{align}
 \label{eq:coex_phi}
 \mu(\phi_0,n_0)&=\mu(\phi_h,n_h),\\
 \label{eq:coex_npm}
 \lambda^\pm(\phi_0,n_0)&=\lambda^\pm(\phi_h,n_h),
\end{align}
where $\phi_h$ and $n_h$ are the composition and salt concentration of the 
water-rich phase, 
respectively.
The chemical potentials are found from:
\begin{align}
\label{eq:mu_bulk}
\mu&=v_0\frac{\partial f_0}{\partial 
\phi}=k_BT\bigl[\log\left(\frac{\phi}{1-\phi}
\right)+\chi(1-2\phi) \nn \\ &-v_0(\Delta
u^+n^++\Delta u^-n^-)\bigr]\\
\label{eq:lam_bulk}
\lambda^\pm&=\frac{\partial f_0 }{\partial n^\pm}=k_BT\bigl[\log \left(v_0 
n^\pm \right)-\Delta u^\pm\phi\bigr]
\end{align}

Using these relations in \cref{eq:coex_phi,eq:coex_npm} we can numerically 
determine $\phi_h$ and $n_h$. The water-rich phase is thermodynamically 
preferable when $\omega_0(\phi_h,n_h)<\omega_0(\phi_0,n_0)$. For a fixed value 
of $n_0$, the locus 
of reservoir compositions $\phi_t(T;n_0,\Delta u^\pm)$ for which
$\omega_0(\phi_h,n_h)=\omega_0(\phi_0,n_0)$ defines a stability diagram in the
$\phi_0-T$ plane. $\phi_t$ is shown for $n_0=0.1$ M,
$\Delta u ^\pm=4$, and point-like ions in \cref{fig_bin} of the main text. At 
$\phi_0=\phi_t$, a so-called precipitation transition \cite{onuki_pre_2010} 
occurs and the composition jumps to $\phi_{h}$ corresponding to $\phi_t$. 

A similar result was obtained by Okamoto and Onuki \cite{onuki_pre_2010} within 
a 
canonical ensemble calculation, and our treatment corresponds to the limit of 
zero volume fraction of the water-rich phase in their work. In our 
work, $\phi_t(T)$ is viewed as the boundary in the $\phi_0-T$ plane for 
which the hydrophilic solute alone can induce a phase change inside the pore. 
In 
the main text, however, we study phase changes inside pores in the range 
$\phi_0<\phi_t(T)$.

\bibliography{refdb}

\begin{thebibliography}{26}%
\makeatletter
\providecommand \@ifxundefined [1]{%
 \@ifx{#1\undefined}
}%
\providecommand \@ifnum [1]{%
 \ifnum #1\expandafter \@firstoftwo
 \else \expandafter \@secondoftwo
 \fi
}%
\providecommand \@ifx [1]{%
 \ifx #1\expandafter \@firstoftwo
 \else \expandafter \@secondoftwo
 \fi
}%
\providecommand \natexlab [1]{#1}%
\providecommand \enquote  [1]{``#1''}%
\providecommand \bibnamefont  [1]{#1}%
\providecommand \bibfnamefont [1]{#1}%
\providecommand \citenamefont [1]{#1}%
\providecommand \href@noop [0]{\@secondoftwo}%
\providecommand \href [0]{\begingroup \@sanitize@url \@href}%
\providecommand \@href[1]{\@@startlink{#1}\@@href}%
\providecommand \@@href[1]{\endgroup#1\@@endlink}%
\providecommand \@sanitize@url [0]{\catcode `\\12\catcode `\$12\catcode
  `\&12\catcode `\#12\catcode `\^12\catcode `\_12\catcode `\%12\relax}%
\providecommand \@@startlink[1]{}%
\providecommand \@@endlink[0]{}%
\providecommand \url  [0]{\begingroup\@sanitize@url \@url }%
\providecommand \@url [1]{\endgroup\@href {#1}{\urlprefix }}%
\providecommand \urlprefix  [0]{URL }%
\providecommand \Eprint [0]{\href }%
\providecommand \doibase [0]{http://dx.doi.org/}%
\providecommand \selectlanguage [0]{\@gobble}%
\providecommand \bibinfo  [0]{\@secondoftwo}%
\providecommand \bibfield  [0]{\@secondoftwo}%
\providecommand \translation [1]{[#1]}%
\providecommand \BibitemOpen [0]{}%
\providecommand \bibitemStop [0]{}%
\providecommand \bibitemNoStop [0]{.\EOS\space}%
\providecommand \EOS [0]{\spacefactor3000\relax}%
\providecommand \BibitemShut  [1]{\csname bibitem#1\endcsname}%
\let\auto@bib@innerbib\@empty
\bibitem [{\citenamefont {Kedem}\ and\ \citenamefont
  {Katchalsky}(1958)}]{katchalsky_bpa_1958}%
  \BibitemOpen
  \bibfield  {author} {\bibinfo {author} {\bibfnamefont {O.}~\bibnamefont
  {Kedem}}\ and\ \bibinfo {author} {\bibfnamefont {A.}~\bibnamefont
  {Katchalsky}},\ }\href {\doibase 10.1016/0006-3002(58)90330-5} {\bibfield
  {journal} {\bibinfo  {journal} {Biochimica et biophysica acta}\ }\textbf
  {\bibinfo {volume} {27}},\ \bibinfo {pages} {229} (\bibinfo {year}
  {1958})}\BibitemShut {NoStop}%
\bibitem [{\citenamefont {Pendergast}\ and\ \citenamefont
  {Hoek}(2011)}]{hoek_ees_2011}%
  \BibitemOpen
  \bibfield  {author} {\bibinfo {author} {\bibfnamefont {M.~M.}\ \bibnamefont
  {Pendergast}}\ and\ \bibinfo {author} {\bibfnamefont {E.~M.~V.}\ \bibnamefont
  {Hoek}},\ }\href {\doibase 10.1039/c0ee00541j} {\bibfield  {journal}
  {\bibinfo  {journal} {Energy \& Environmental Science}\ }\textbf {\bibinfo
  {volume} {4}},\ \bibinfo {pages} {1946} (\bibinfo {year} {2011})}\BibitemShut
  {NoStop}%
\bibitem [{\citenamefont {Carta}\ \emph {et~al.}(2013)\citenamefont {Carta},
  \citenamefont {Malpass-Evans}, \citenamefont {Croad}, \citenamefont {Rogan},
  \citenamefont {Jansen}, \citenamefont {Bernardo}, \citenamefont
  {Bazzarelli},\ and\ \citenamefont {McKeown}}]{neil_science_2013}%
  \BibitemOpen
  \bibfield  {author} {\bibinfo {author} {\bibfnamefont {M.}~\bibnamefont
  {Carta}}, \bibinfo {author} {\bibfnamefont {R.}~\bibnamefont
  {Malpass-Evans}}, \bibinfo {author} {\bibfnamefont {M.}~\bibnamefont
  {Croad}}, \bibinfo {author} {\bibfnamefont {Y.}~\bibnamefont {Rogan}},
  \bibinfo {author} {\bibfnamefont {J.~C.}\ \bibnamefont {Jansen}}, \bibinfo
  {author} {\bibfnamefont {P.}~\bibnamefont {Bernardo}}, \bibinfo {author}
  {\bibfnamefont {F.}~\bibnamefont {Bazzarelli}}, \ and\ \bibinfo {author}
  {\bibfnamefont {N.~B.}\ \bibnamefont {McKeown}},\ }\href {\doibase
  10.1126/science.1228032} {\bibfield  {journal} {\bibinfo  {journal}
  {Science}\ }\textbf {\bibinfo {volume} {339}},\ \bibinfo {pages} {303}
  (\bibinfo {year} {2013})}\BibitemShut {NoStop}%
\bibitem [{\citenamefont {Yang}\ and\ \citenamefont
  {Yang}(2003)}]{yang_jms_2003}%
  \BibitemOpen
  \bibfield  {author} {\bibinfo {author} {\bibfnamefont {B.}~\bibnamefont
  {Yang}}\ and\ \bibinfo {author} {\bibfnamefont {W.~T.}\ \bibnamefont
  {Yang}},\ }\href {\doibase 10.1016/s0376-7388(03)00182-0} {\bibfield
  {journal} {\bibinfo  {journal} {Journal of Membrane Science}\ }\textbf
  {\bibinfo {volume} {218}},\ \bibinfo {pages} {247} (\bibinfo {year}
  {2003})},\ \bibinfo {note} {86}\BibitemShut {NoStop}%
\bibitem [{\citenamefont {Yameen}\ \emph {et~al.}(2009)\citenamefont {Yameen},
  \citenamefont {Ali}, \citenamefont {Neumann}, \citenamefont {Ensinger},
  \citenamefont {Knoll},\ and\ \citenamefont {Azzaroni}}]{azzaroni_small_2009}%
  \BibitemOpen
  \bibfield  {author} {\bibinfo {author} {\bibfnamefont {B.}~\bibnamefont
  {Yameen}}, \bibinfo {author} {\bibfnamefont {M.}~\bibnamefont {Ali}},
  \bibinfo {author} {\bibfnamefont {R.}~\bibnamefont {Neumann}}, \bibinfo
  {author} {\bibfnamefont {W.}~\bibnamefont {Ensinger}}, \bibinfo {author}
  {\bibfnamefont {W.}~\bibnamefont {Knoll}}, \ and\ \bibinfo {author}
  {\bibfnamefont {O.}~\bibnamefont {Azzaroni}},\ }\href {\doibase
  10.1002/smll.200801318} {\bibfield  {journal} {\bibinfo  {journal} {Small}\
  }\textbf {\bibinfo {volume} {5}},\ \bibinfo {pages} {1287} (\bibinfo {year}
  {2009})},\ \bibinfo {note} {87}\BibitemShut {NoStop}%
\bibitem [{\citenamefont {Hou}\ \emph {et~al.}(2015)\citenamefont {Hou},
  \citenamefont {Hu}, \citenamefont {Grinthal}, \citenamefont {Khan},\ and\
  \citenamefont {Aizenberg}}]{aizenberg_nature_2015}%
  \BibitemOpen
  \bibfield  {author} {\bibinfo {author} {\bibfnamefont {X.}~\bibnamefont
  {Hou}}, \bibinfo {author} {\bibfnamefont {Y.}~\bibnamefont {Hu}}, \bibinfo
  {author} {\bibfnamefont {A.}~\bibnamefont {Grinthal}}, \bibinfo {author}
  {\bibfnamefont {M.}~\bibnamefont {Khan}}, \ and\ \bibinfo {author}
  {\bibfnamefont {J.}~\bibnamefont {Aizenberg}},\ }\href {\doibase
  10.1038/nature14253} {\bibfield  {journal} {\bibinfo  {journal} {Nature}\
  }\textbf {\bibinfo {volume} {519}},\ \bibinfo {pages} {70} (\bibinfo {year}
  {2015})},\ \bibinfo {note} {10}\BibitemShut {NoStop}%
\bibitem [{\citenamefont {Dzubiella}\ \emph {et~al.}(2004)\citenamefont
  {Dzubiella}, \citenamefont {Allen},\ and\ \citenamefont
  {Hansen}}]{hansen_jcp_2004}%
  \BibitemOpen
  \bibfield  {author} {\bibinfo {author} {\bibfnamefont {J.}~\bibnamefont
  {Dzubiella}}, \bibinfo {author} {\bibfnamefont {R.~J.}\ \bibnamefont
  {Allen}}, \ and\ \bibinfo {author} {\bibfnamefont {J.~P.}\ \bibnamefont
  {Hansen}},\ }\href {\doibase 10.1063/1.1665656} {\bibfield  {journal}
  {\bibinfo  {journal} {Journal of Chemical Physics}\ }\textbf {\bibinfo
  {volume} {120}},\ \bibinfo {pages} {5001} (\bibinfo {year}
  {2004})}\BibitemShut {NoStop}%
\bibitem [{\citenamefont {Dzubiella}\ and\ \citenamefont
  {Hansen}(2005)}]{hansen_jcp_2005}%
  \BibitemOpen
  \bibfield  {author} {\bibinfo {author} {\bibfnamefont {J.}~\bibnamefont
  {Dzubiella}}\ and\ \bibinfo {author} {\bibfnamefont {J.~P.}\ \bibnamefont
  {Hansen}},\ }\href {\doibase 10.1063/1.1927514} {\bibfield  {journal}
  {\bibinfo  {journal} {Journal of Chemical Physics}\ }\textbf {\bibinfo
  {volume} {122}} (\bibinfo {year} {2005}),\ 10.1063/1.1927514}\BibitemShut
  {NoStop}%
\bibitem [{\citenamefont {Powell}\ \emph {et~al.}(2011)\citenamefont {Powell},
  \citenamefont {Cleary}, \citenamefont {Davenport}, \citenamefont {Shea},\
  and\ \citenamefont {Siwy}}]{siwy_nature_nanotech_2011}%
  \BibitemOpen
  \bibfield  {author} {\bibinfo {author} {\bibfnamefont {M.~R.}\ \bibnamefont
  {Powell}}, \bibinfo {author} {\bibfnamefont {L.}~\bibnamefont {Cleary}},
  \bibinfo {author} {\bibfnamefont {M.}~\bibnamefont {Davenport}}, \bibinfo
  {author} {\bibfnamefont {K.~J.}\ \bibnamefont {Shea}}, \ and\ \bibinfo
  {author} {\bibfnamefont {Z.~S.}\ \bibnamefont {Siwy}},\ }\href {\doibase
  10.1038/nnano.2011.189} {\bibfield  {journal} {\bibinfo  {journal} {Nature
  Nanotechnology}\ }\textbf {\bibinfo {volume} {6}},\ \bibinfo {pages} {798}
  (\bibinfo {year} {2011})},\ \bibinfo {note} {81}\BibitemShut {NoStop}%
\bibitem [{\citenamefont {Smirnov}\ \emph {et~al.}(2011)\citenamefont
  {Smirnov}, \citenamefont {Vlassiouk},\ and\ \citenamefont
  {Lavrik}}]{lavrik_acsnano_2011}%
  \BibitemOpen
  \bibfield  {author} {\bibinfo {author} {\bibfnamefont {S.~N.}\ \bibnamefont
  {Smirnov}}, \bibinfo {author} {\bibfnamefont {I.~V.}\ \bibnamefont
  {Vlassiouk}}, \ and\ \bibinfo {author} {\bibfnamefont {N.~V.}\ \bibnamefont
  {Lavrik}},\ }\href {\doibase 10.1021/nn202392d} {\bibfield  {journal}
  {\bibinfo  {journal} {Acs Nano}\ }\textbf {\bibinfo {volume} {5}},\ \bibinfo
  {pages} {7453} (\bibinfo {year} {2011})}\BibitemShut {NoStop}%
\bibitem [{\citenamefont {Tsori}\ and\ \citenamefont
  {Leibler}(2007)}]{tsori_pnas_2007}%
  \BibitemOpen
  \bibfield  {author} {\bibinfo {author} {\bibfnamefont {Y.}~\bibnamefont
  {Tsori}}\ and\ \bibinfo {author} {\bibfnamefont {L.}~\bibnamefont
  {Leibler}},\ }\href {\doibase 10.1073/pnas.0607746104} {\bibfield  {journal}
  {\bibinfo  {journal} {Proc. Nat. Acad. Sci.}\ }\textbf {\bibinfo {volume}
  {104}},\ \bibinfo {pages} {7348} (\bibinfo {year} {2007})}\BibitemShut
  {NoStop}%
\bibitem [{\citenamefont {Okamoto}\ and\ \citenamefont
  {Onuki}(2010)}]{onuki_pre_2010}%
  \BibitemOpen
  \bibfield  {author} {\bibinfo {author} {\bibfnamefont {R.}~\bibnamefont
  {Okamoto}}\ and\ \bibinfo {author} {\bibfnamefont {A.}~\bibnamefont
  {Onuki}},\ }\href {\doibase 10.1103/PhysRevE.82.051501} {\bibfield  {journal}
  {\bibinfo  {journal} {Phys. Rev. E}\ }\textbf {\bibinfo {volume} {82}},\
  \bibinfo {pages} {051501} (\bibinfo {year} {2010})}\BibitemShut {NoStop}%
\bibitem [{\citenamefont {Bier}\ \emph {et~al.}(2012)\citenamefont {Bier},
  \citenamefont {Gambassi},\ and\ \citenamefont {Dietrich}}]{bier2012}%
  \BibitemOpen
  \bibfield  {author} {\bibinfo {author} {\bibfnamefont {M.}~\bibnamefont
  {Bier}}, \bibinfo {author} {\bibfnamefont {A.}~\bibnamefont {Gambassi}}, \
  and\ \bibinfo {author} {\bibfnamefont {S.}~\bibnamefont {Dietrich}},\ }\href
  {\doibase 10.1063/1.4733973} {\bibfield  {journal} {\bibinfo  {journal} {J.
  Chem. Phys.}\ }\textbf {\bibinfo {volume} {137}},\ \bibinfo {eid} {034504}
  (\bibinfo {year} {2012})}\BibitemShut {NoStop}%
\bibitem [{\citenamefont {Leunissen}\ \emph {et~al.}(2007)\citenamefont
  {Leunissen}, \citenamefont {van Blaaderen}, \citenamefont {Hollingsworth},
  \citenamefont {Sullivan},\ and\ \citenamefont {Chaikin}}]{leunissen2007}%
  \BibitemOpen
  \bibfield  {author} {\bibinfo {author} {\bibfnamefont {M.~E.}\ \bibnamefont
  {Leunissen}}, \bibinfo {author} {\bibfnamefont {A.}~\bibnamefont {van
  Blaaderen}}, \bibinfo {author} {\bibfnamefont {A.~D.}\ \bibnamefont
  {Hollingsworth}}, \bibinfo {author} {\bibfnamefont {M.~T.}\ \bibnamefont
  {Sullivan}}, \ and\ \bibinfo {author} {\bibfnamefont {P.~M.}\ \bibnamefont
  {Chaikin}},\ }\href {\doibase 10.1073/pnas.0610589104} {\bibfield  {journal}
  {\bibinfo  {journal} {Proc. Natl. Acad. Sci. U.S.A.}\ }\textbf {\bibinfo
  {volume} {104}},\ \bibinfo {pages} {2585} (\bibinfo {year}
  {2007})}\BibitemShut {NoStop}%
\bibitem [{\citenamefont {Nellen}\ \emph {et~al.}(2011)\citenamefont {Nellen},
  \citenamefont {Dietrich}, \citenamefont {Helden}, \citenamefont {Chodankar},
  \citenamefont {Nyg\aa{}rd}, \citenamefont {van~der Veen},\ and\ \citenamefont
  {Bechinger}}]{nellen2011}%
  \BibitemOpen
  \bibfield  {author} {\bibinfo {author} {\bibfnamefont {U.}~\bibnamefont
  {Nellen}}, \bibinfo {author} {\bibfnamefont {J.}~\bibnamefont {Dietrich}},
  \bibinfo {author} {\bibfnamefont {L.}~\bibnamefont {Helden}}, \bibinfo
  {author} {\bibfnamefont {S.}~\bibnamefont {Chodankar}}, \bibinfo {author}
  {\bibfnamefont {K.}~\bibnamefont {Nyg\aa{}rd}}, \bibinfo {author}
  {\bibfnamefont {J.~F.}\ \bibnamefont {van~der Veen}}, \ and\ \bibinfo
  {author} {\bibfnamefont {C.}~\bibnamefont {Bechinger}},\ }\href@noop {}
  {\bibfield  {journal} {\bibinfo  {journal} {Soft Matter}\ }\textbf {\bibinfo
  {volume} {7}},\ \bibinfo {pages} {5360} (\bibinfo {year} {2011})}\BibitemShut
  {NoStop}%
\bibitem [{\citenamefont {Samin}\ \emph {et~al.}(2014)\citenamefont {Samin},
  \citenamefont {Hod}, \citenamefont {Melamed}, \citenamefont {Gottlieb},\ and\
  \citenamefont {Tsori}}]{samin2014}%
  \BibitemOpen
  \bibfield  {author} {\bibinfo {author} {\bibfnamefont {S.}~\bibnamefont
  {Samin}}, \bibinfo {author} {\bibfnamefont {M.}~\bibnamefont {Hod}}, \bibinfo
  {author} {\bibfnamefont {E.}~\bibnamefont {Melamed}}, \bibinfo {author}
  {\bibfnamefont {M.}~\bibnamefont {Gottlieb}}, \ and\ \bibinfo {author}
  {\bibfnamefont {Y.}~\bibnamefont {Tsori}},\ }\href {\doibase
  10.1103/physrevapplied.2.024008} {\bibfield  {journal} {\bibinfo  {journal}
  {Physical Review Applied}\ }\textbf {\bibinfo {volume} {2}} (\bibinfo {year}
  {2014}),\ 10.1103/physrevapplied.2.024008}\BibitemShut {NoStop}%
\bibitem [{\citenamefont {Samin}\ and\ \citenamefont
  {Tsori}(2011)}]{efips_epl}%
  \BibitemOpen
  \bibfield  {author} {\bibinfo {author} {\bibfnamefont {S.}~\bibnamefont
  {Samin}}\ and\ \bibinfo {author} {\bibfnamefont {Y.}~\bibnamefont {Tsori}},\
  }\href@noop {} {\bibfield  {journal} {\bibinfo  {journal} {EPL}\ }\textbf
  {\bibinfo {volume} {95}},\ \bibinfo {pages} {36002} (\bibinfo {year}
  {2011})}\BibitemShut {NoStop}%
\bibitem [{\citenamefont {Okamoto}\ and\ \citenamefont
  {Onuki}(2011)}]{onuki_pre_2011}%
  \BibitemOpen
  \bibfield  {author} {\bibinfo {author} {\bibfnamefont {R.}~\bibnamefont
  {Okamoto}}\ and\ \bibinfo {author} {\bibfnamefont {A.}~\bibnamefont
  {Onuki}},\ }\href@noop {} {\bibfield  {journal} {\bibinfo  {journal} {Phys.
  Rev. E}\ }\textbf {\bibinfo {volume} {84}},\ \bibinfo {pages} {051401}
  (\bibinfo {year} {2011})}\BibitemShut {NoStop}%
\bibitem [{\citenamefont {Samin}\ and\ \citenamefont
  {Tsori}(2012)}]{efips_jcp_2012}%
  \BibitemOpen
  \bibfield  {author} {\bibinfo {author} {\bibfnamefont {S.}~\bibnamefont
  {Samin}}\ and\ \bibinfo {author} {\bibfnamefont {Y.}~\bibnamefont {Tsori}},\
  }\href@noop {} {\bibfield  {journal} {\bibinfo  {journal} {J. Chem. Phys.}\
  }\textbf {\bibinfo {volume} {136}},\ \bibinfo {eid} {154908} (\bibinfo {year}
  {2012})}\BibitemShut {NoStop}%
\bibitem [{\citenamefont {Pousaneh}\ and\ \citenamefont
  {Ciach}(2014)}]{pousaneh2014}%
  \BibitemOpen
  \bibfield  {author} {\bibinfo {author} {\bibfnamefont {F.}~\bibnamefont
  {Pousaneh}}\ and\ \bibinfo {author} {\bibfnamefont {A.}~\bibnamefont
  {Ciach}},\ }\href {\doibase 10.1039/C4SM01264J} {\bibfield  {journal}
  {\bibinfo  {journal} {Soft Matter}\ }\textbf {\bibinfo {volume} {10}},\
  \bibinfo {pages} {8188} (\bibinfo {year} {2014})}\BibitemShut {NoStop}%
\bibitem [{\citenamefont {Borukhov}\ \emph {et~al.}(1997)\citenamefont
  {Borukhov}, \citenamefont {Andelman},\ and\ \citenamefont
  {Orland}}]{borukhov1997}%
  \BibitemOpen
  \bibfield  {author} {\bibinfo {author} {\bibfnamefont {I.}~\bibnamefont
  {Borukhov}}, \bibinfo {author} {\bibfnamefont {D.}~\bibnamefont {Andelman}},
  \ and\ \bibinfo {author} {\bibfnamefont {H.}~\bibnamefont {Orland}},\
  }\href@noop {} {\bibfield  {journal} {\bibinfo  {journal} {Phys. Rev. Lett.}\
  }\textbf {\bibinfo {volume} {79}},\ \bibinfo {pages} {435} (\bibinfo {year}
  {1997})}\BibitemShut {NoStop}%
\bibitem [{\citenamefont {Onuki}\ and\ \citenamefont
  {Kitamura}(2004)}]{onuki_jcp_2004}%
  \BibitemOpen
  \bibfield  {author} {\bibinfo {author} {\bibfnamefont {A.}~\bibnamefont
  {Onuki}}\ and\ \bibinfo {author} {\bibfnamefont {H.}~\bibnamefont
  {Kitamura}},\ }\href {\doibase 10.1063/1.1769357} {\bibfield  {journal}
  {\bibinfo  {journal} {J. Chem. Phys.}\ }\textbf {\bibinfo {volume} {121}},\
  \bibinfo {pages} {3143} (\bibinfo {year} {2004})}\BibitemShut {NoStop}%
\bibitem [{\citenamefont {Kalidas}\ \emph {et~al.}(2000)\citenamefont
  {Kalidas}, \citenamefont {Hefter},\ and\ \citenamefont
  {Marcus}}]{marcus_cation}%
  \BibitemOpen
  \bibfield  {author} {\bibinfo {author} {\bibfnamefont {C.}~\bibnamefont
  {Kalidas}}, \bibinfo {author} {\bibfnamefont {G.}~\bibnamefont {Hefter}}, \
  and\ \bibinfo {author} {\bibfnamefont {Y.}~\bibnamefont {Marcus}},\ }\href
  {\doibase 10.1021/cr980144k} {\bibfield  {journal} {\bibinfo  {journal}
  {Chem. Rev.}\ }\textbf {\bibinfo {volume} {100}},\ \bibinfo {pages} {819}
  (\bibinfo {year} {2000})}\BibitemShut {NoStop}%
\bibitem [{\citenamefont {Marcus}(2007)}]{marcus_anion}%
  \BibitemOpen
  \bibfield  {author} {\bibinfo {author} {\bibfnamefont {Y.}~\bibnamefont
  {Marcus}},\ }\href {\doibase 10.1021/cr068045r} {\bibfield  {journal}
  {\bibinfo  {journal} {Chem. Rev.}\ }\textbf {\bibinfo {volume} {107}},\
  \bibinfo {pages} {3880} (\bibinfo {year} {2007})}\BibitemShut {NoStop}%
\bibitem [{\citenamefont {Tsori}(2009)}]{tsori_rmp_2009}%
  \BibitemOpen
  \bibfield  {author} {\bibinfo {author} {\bibfnamefont {Y.}~\bibnamefont
  {Tsori}},\ }\href {\doibase 10.1103/RevModPhys.81.1471} {\bibfield  {journal}
  {\bibinfo  {journal} {Reviews of Modern Physics}\ }\textbf {\bibinfo {volume}
  {81}},\ \bibinfo {pages} {1471} (\bibinfo {year} {2009})}\BibitemShut
  {NoStop}%
\bibitem [{\citenamefont {Ben-Yaakov}\ \emph {et~al.}(2009)\citenamefont
  {Ben-Yaakov}, \citenamefont {Andelman}, \citenamefont {Harries},\ and\
  \citenamefont {Podgornik}}]{andelman_jpcb_2009}%
  \BibitemOpen
  \bibfield  {author} {\bibinfo {author} {\bibfnamefont {D.}~\bibnamefont
  {Ben-Yaakov}}, \bibinfo {author} {\bibfnamefont {D.}~\bibnamefont
  {Andelman}}, \bibinfo {author} {\bibfnamefont {D.}~\bibnamefont {Harries}}, \
  and\ \bibinfo {author} {\bibfnamefont {R.}~\bibnamefont {Podgornik}},\
  }\href@noop {} {\bibfield  {journal} {\bibinfo  {journal} {J. Phys. Chem. B}\
  }\textbf {\bibinfo {volume} {113}},\ \bibinfo {pages} {6001} (\bibinfo {year}
  {2009})}\BibitemShut {NoStop}%
\end{thebibliography}%

\end{document}